\documentclass{optica-article}

\journal{opticajournal} 

\articletype{Research Article}

\usepackage{placeins}

\newcommand{\safeincludegraphics}[2][]{%
  \IfFileExists{#2}{%
    \includegraphics[#1]{#2}%
  }{%
    \fbox{\parbox[c][0.22\textheight][c]{0.9\linewidth}{\centering Missing figure asset\\\texttt{#2}}}%
  }%
}

\begin{document}

\title{Scalable Photonic Neural Networks via Surrogate Scattering-Matrix Inverse Design}

\author{Azka Maula Iskandar Muda\authormark{1} and U\u{g}ur Te\u{g}in\authormark{1,*}}

\address{\authormark{1}Department of Electrical and Electronics Engineering, Ko\c{c} University, \.{I}stanbul, T\"{u}rkiye}

\email{\authormark{*}utegin@ku.edu.tr}


\begin{abstract*}
Inverse-designed nanophotonic media are a promising platform for compact optical neural networks, but training them end to end is expensive because each adjoint iteration couples the full-wave solver to the dataset minibatch, so the number of electromagnetic simulations scales with both the network depth and the batch size. We introduce a two-stage surrogate workflow that decouples task learning from electromagnetic realization. In the first stage, the trainable optical block is represented as a passive complex matrix with bounded singular values and the classification task is solved directly in matrix space at negligible cost. In the second stage, the selected target operator is transferred to a fabrication-aware freeform device through an adjoint problem driven by a Frobenius-norm transmission residual and a reflection penalty, which removes the minibatch dependence from the full-wave loop and yields a smoother loss landscape than intensity-domain cross-entropy. We further introduce a banded-router architecture composed with a fixed evanescent-coupling region, which exploits the bandwidth-additive property of matrix products to realize dense effective operators within a design region roughly half as long as a fully local router would require. The framework is validated on three tasks. On MedMNIST, the realized all-optical classifier reproduces the surrogate accuracy within $0.6$ percentage points after only 20 adjoint epochs. On RSSCN7, the banded router plus evanescent stage improves test accuracy by more than 15 percentage points over a linear readout baseline. A Yin-Yang task confirms that the same framework supports nonlinear decision boundaries. These results indicate that surrogate-guided inverse design is a practical route to training compact photonic processors with simulation budgets orders of magnitude smaller than direct geometry-to-task pipelines.
\end{abstract*}


\section{Introduction}
Deep neural networks now drive a broad range of applications, from image recognition to natural language processing. Training these models demands growing amounts of compute, and digital electronics is approaching its efficiency limits. Optical computing offers a hardware route to address this bottleneck. Photonic neural networks perform matrix-vector multiplication at the speed of light with passive components. Several architectures have been demonstrated, including coherent Mach-Zehnder meshes \cite{Shen2017DeepLearning,Hughes2018Adjoint,Pai2023Experimental}, diffractive deep networks \cite{Lin2018DONN}, phase-change photonic tensor cores \cite{Feldmann2021Parallel}, and on-chip photonic classifiers \cite{Ashtiani2022Chip}. Recent reviews frame these systems as physical neural networks whose capacity is set by the interplay between trainable parameters and hardware physics \cite{Wetzstein2020Inference,Liu2024ONNReview,Momeni2025PNN,Xue2024FFM}.

Inverse-designed nanophotonic media offer the most compact route to optical linear operators. Instead of cascading discrete building blocks, the entire passive device is treated as a continuous permittivity distribution and optimized so that its scattering matrix implements the desired map \cite{Molesky2018InverseDesign,Piggott2017FabricationConstrained}. Gradients are computed with the adjoint method \cite{LalauKeraly2013Adjoint}. Khoram \emph{et al.} first showed that freeform nanophotonic media can perform neural inference \cite{Khoram2019NanophotonicMedia}. Nikkhah \emph{et al.} experimentally realized vector-matrix multiplication in low-index-contrast structures \cite{Nikkhah2024VectorMatrix}. Zhao \emph{et al.} and Sved \emph{et al.} have since pushed the computational density and scale of these devices \cite{Zhao2025HighDensityMedia,Sved2026PNNAccelerators}. Fabrication-aware maps, such as conic filtering and subpixel-smoothed projection \cite{Hammond2025SSP}, allow these designs to be binarized under minimum-feature-size constraints.

Training an inverse-designed photonic classifier end to end remains expensive. In a geometry-to-task formulation, each adjoint iteration calls the full-wave solver for every input in a minibatch. The number of electromagnetic simulations scales as $O(N_{\mathrm{fwd}}\cdot N_{\mathrm{bwd}}\cdot B)$ per epoch, where $B$ is the batch size. This coupling dominates the training cost and limits the number of epochs one can afford. It also mixes three different sources of nonconvexity in a single loop: label fitting, fabrication-aware parameterization, and electromagnetic scattering.

Here we decouple these roles with a two-stage surrogate workflow. In the first stage, the trainable optical block is optimized directly in matrix space using a singular-value-bounded parameterization, so task learning can be carried out at negligible cost. The resulting operator $T^{\star}$ is then transferred to a fabrication-aware freeform device through an adjoint problem. The realization objective is a Frobenius-norm residual between the simulated transmission matrix and $T^{\star}$, together with a reflection penalty. Because $T^{\star}$ is fixed, the full-wave stage is batch-free. Its cost scales as $O(N_{\mathrm{fwd}}\cdot N_{\mathrm{bwd}})$ per epoch, independent of dataset size. The loss also specifies the required amplitude and phase at every output port, which yields smoother convergence than intensity-domain cross-entropy.

Our second contribution addresses a layout bottleneck in inverse-designed routers. When the target operator routes power from one edge port to the opposite edge, the design region must be long. We impose locality as an architectural prior by constraining the trainable operator to a banded matrix with half-bandwidth $w$. We recover global mixing by composing it with a fixed downstream evanescent-coupling region. Since the product of two banded matrices has half-bandwidth at most $w_A+w_B$, a dense effective operator can emerge from two individually sparse factors \cite{HornJohnson2012MatrixAnalysis,Dao2019Butterfly}. This halves the required propagation length of the freeform region and proportionally reduces the number of design grid points.

\begin{figure}[htbp]
\centering
\includegraphics[width=\linewidth]{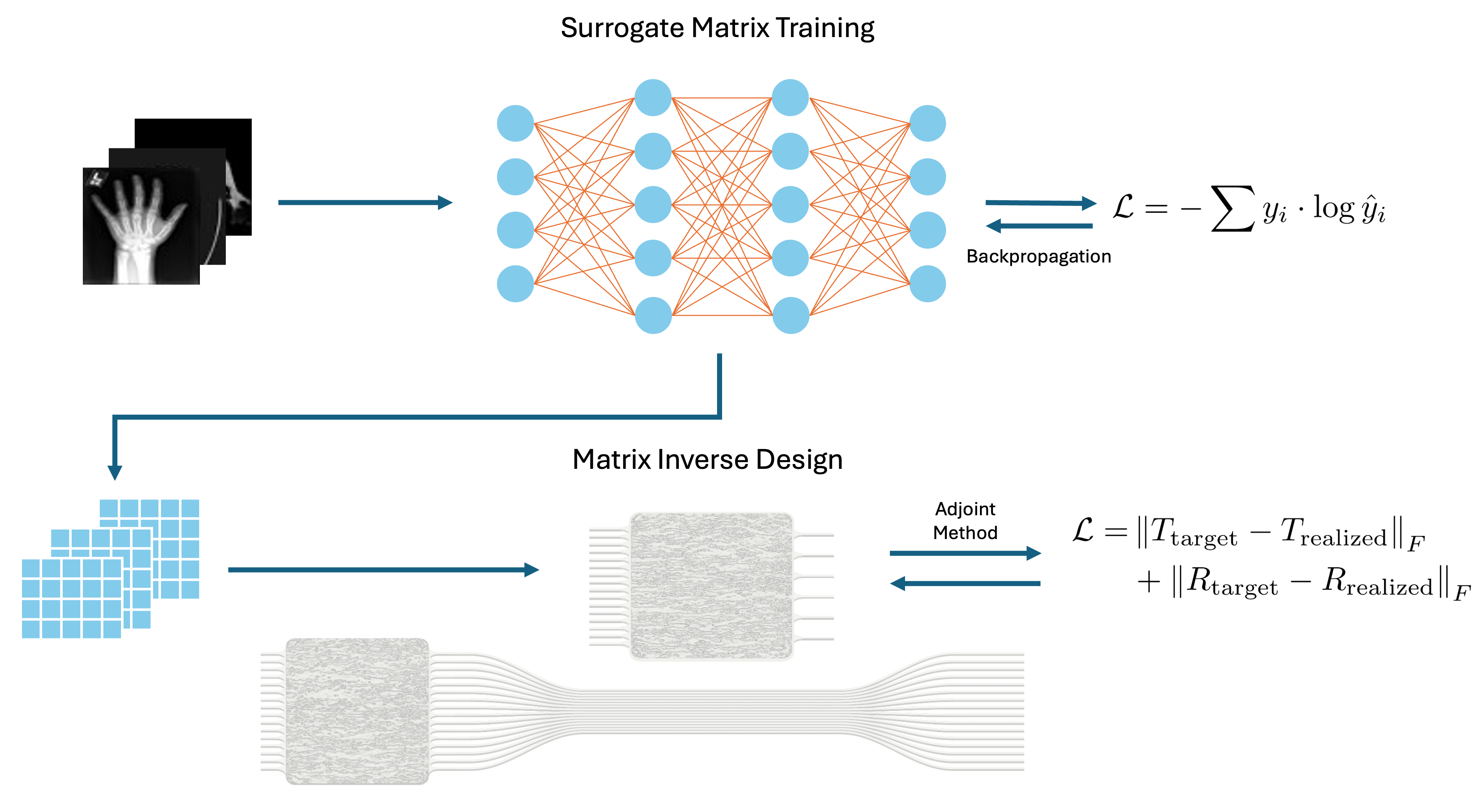}
\caption{Surrogate matrix-based inverse design of an optical neural network. In the surrogate stage (top), a passive complex matrix is trained directly on the classification task with cross-entropy loss. In the realization stage (bottom), the adjoint method is used to match the transmission and reflection blocks of a fabrication-aware freeform device to the surrogate target via a Frobenius-norm residual.}
\label{fig:overview}
\end{figure}

We validate the framework on three tasks. For MedMNIST \cite{Yang2023MedMNISTv2}, the realized all-optical classifier matches the surrogate accuracy within $0.6$ percentage points after 20 adjoint epochs, an order of magnitude fewer than comparable geometry-to-task pipelines \cite{Sved2026PNNAccelerators}. For RSSCN7 \cite{Zou2015RSSCN7}, the chained banded-router plus evanescent-stage architecture improves on a linear readout baseline by more than 15 percentage points. A Yin-Yang task shows that the same framework supports nonlinear decision boundaries with intensity detection and a small digital readout. These results establish surrogate-guided inverse design as a practical route to training compact photonic processors with simulation budgets orders of magnitude smaller than direct geometry-to-task pipelines. A schematic of the workflow is shown in Fig.~\ref{fig:overview}.

\section{Method}
The workflow has two stages. First, we solve the learning task in a low-cost surrogate space by representing the trainable optical block directly as a passive matrix. We then transfer the selected surrogate operator to a fabrication-aware nanophotonic device through adjoint inverse design. This separation removes dataset batching from the expensive electromagnetic loop and lets the full-wave solver focus on matching a target scattering response.

\subsection{Surrogate operator model}
Let
$$
\mathcal{D}=\{(z_b,y_b)\}_{b=1}^{B}
$$
denote the dataset. We first map each sample \(z\) to a reduced feature vector through dimensionality reductions
$$
x=P(z)\in\mathbb{R}^{N_{\mathrm{in}}},
$$
and then optically encode it as
$$
a=\phi(x)\in\mathbb{C}^{N_{\mathrm{in}}}.
$$
The local trainable optical stage is modeled as a passive linear operator
$$
M\in\mathbb{C}^{N_{\mathrm{out}}\times N_{\mathrm{in}}},
$$
parameterized by
$$
M=U\,\Sigma(s)\,V^{\dagger}.
$$
Here \(V\in\mathbb{C}^{N_{\mathrm{in}}\times N_{\mathrm{in}}}\) and \(U\in\mathbb{C}^{N_{\mathrm{out}}\times N_{\mathrm{out}}}\) are unitary matrices acting on the input and output spaces, respectively. We generate these factors from learnable skew-Hermitian parameters, which preserves unitarity throughout optimization. The matrix \(\Sigma(s)\in\mathbb{R}^{N_{\mathrm{out}}\times N_{\mathrm{in}}}\) is rectangular. Its \(r=\min(N_{\mathrm{out}},N_{\mathrm{in}})\) principal-diagonal entries are collected in the real singular-value vector \(s\in\mathbb{R}^{r}\). In this factorization, \(U\) and \(V\) represent basis changes, whereas \(s\) captures channel-wise attenuation. We constrain the singular values elementwise as \(s_i\in[s_{\min},s_{\max}]\) with \(0\le s_{\min}\le s_{\max}\le 1\), which directly enforces passivity and insertion-loss bounds.

A fixed downstream optical stage with compatible dimensions is represented by a linear map \(O\). The optical field after the passive stack is
$$
u=OMa,
$$
and the measured optical features are given by the componentwise readout map
$$
\Psi(u)=|u|^2.
$$
Together, these definitions cover both all-optical and cascaded classifiers. For an all-optical system, \(O=I\). In chained systems, \(O\) denotes a fixed downstream photonic operator, such as the evanescent-coupling stage used in RSSCN7.

Surrogate training optimizes the passive operator through
$$
\min_{U,V,s}
\;
\frac{1}{B}\sum_{b=1}^{B}
\mathcal{L}_{\mathrm{task}}\!\big(\Psi(OMa_b),y_b\big)
\;+\;
\sum_k \lambda_k \mathcal{R}_k(U,V,s),
$$
where \(M=U\Sigma(s)V^{\dagger}\) and \(\mathcal{R}_k\) denote optional regularizers on the matrix or on its factors. Because this stage involves only matrix products and elementary nonlinearities, the long exploratory phase of optimization can be carried out without electromagnetic simulation.

We consider two task formulations. For all-optical classification, the optical intensities themselves serve as class logits,
$$
\ell_{\mathrm{AO}}(M)
=
-\frac{1}{B}\sum_{b=1}^{B}
\log\frac{\exp(I_{b,y_b})}{\sum_c \exp(I_{b,c})},
\qquad
I_b=|OMa_b|^2.
$$
For readout-based classification, the optical device acts as a feature extractor, and a linear head is applied to the detected intensities. If \(I_b=|OMa_b|^2\), then the logits are
$$
g_b=W^\top I_b+b,
$$
followed by the same cross-entropy loss. Throughout this work, MedMNIST is treated primarily as an all-optical problem, whereas RSSCN7 is evaluated after composition with a fixed downstream optical stage and a linear readout.

From a computational standpoint, the advantage of the surrogate stage is straightforward. If \(C_{\mathrm{LA}}\) denotes the cost of one surrogate update and \(C_{\mathrm{EM}}\) the cost of one electromagnetic realization step, then typically \(C_{\mathrm{LA}}\ll C_{\mathrm{EM}}\). The two-stage workflow therefore assigns the long exploratory search to the inexpensive surrogate regime and reserves full-wave optimization for a shorter operator-matching stage. Once a target operator \(T^\star\) is fixed, the expensive stage becomes a residual-matching problem in scattering space rather than a label-composed learning problem \cite{Khoram2019NanophotonicMedia,Zhao2025HighDensityMedia,Sved2026PNNAccelerators}.

\subsection{Fabrication-aware inverse design}
Once surrogate training is complete, the selected operator \(M^\star\) defines the local transmission target,
$$
T^\star=M^\star.
$$
During inverse design, only the freeform design region is optimized; the input and output waveguides, tapers, and any optional fixed evanescent-coupling region remain fixed. We denote the latent design variables by
$$
\theta_{\mathrm{raw}}\in[0,1]^{H\times W},
$$
and pass through a fabrication-aware map before simulation,
$$
\theta_{\mathrm{filt}}=F_{r_f}(\theta_{\mathrm{raw}}),
\qquad
\theta_{\mathrm{sim}}=P_{\beta,\eta,\rho}(\theta_{\mathrm{filt}}).
$$
Within this map, \(F_{r_f}\) is a conic filter that enforces a minimum feature size, and \(P_{\beta,\eta,\rho}\) is a projection operator that drives the design toward a binary pattern while preserving differentiability. We use the subpixel-smoothed projection of Hammond \emph{et al.}~\cite{Hammond2025SSP} to enforce differentiable binarization.

We obtain the simulated permittivity from the projected design variables through
$$
\varepsilon(\theta_{\mathrm{sim}})
=
\varepsilon_{\min}+\theta_{\mathrm{sim}}(\varepsilon_{\max}-\varepsilon_{\min}),
\qquad
\varepsilon_{\min}=\left(n_{\mathrm{eff}}^{150\,\mathrm{nm}}\right)^2,
\quad
\varepsilon_{\max}=\left(n_{\mathrm{eff}}^{220\,\mathrm{nm}}\right)^2.
$$
Electromagnetic simulations are carried out in Tidy3D using a two-dimensional effective-index model at \(\lambda=1.55\,\mu\mathrm{m}\). It captures the relevant in-plane scattering and mode conversion while following an effective-index approximation that has been validated experimentally for similar slab-waveguide structures \cite{Nikkhah2024VectorMatrix}.

Let
$$
T(\theta_{\mathrm{sim}})\in\mathbb{C}^{N_{\mathrm{out}}\times N_{\mathrm{in}}},
\qquad
R(\theta_{\mathrm{sim}})\in\mathbb{C}^{N_{\mathrm{in}}\times N_{\mathrm{in}}}
$$
denote the transmission and reflection blocks extracted from simulation. We then define the realization objective as
$$
\mathcal{L}_{\mathrm{realize}}(\theta_{\mathrm{raw}})
=
w_T\|T(\theta_{\mathrm{sim}})-T^\star\|_F
+
w_R\|R(\theta_{\mathrm{sim}})\|_F
+
\sum_j \mu_j \mathcal{P}_j(\theta_{\mathrm{sim}}).
$$
It matches the desired local transmission while suppressing reflection, without directly differentiating the end-to-end task loss of the full optical classifier.

By comparison, a direct geometry-to-task formulation would optimize
$$
\mathcal{L}_{\mathrm{task}}(\theta_{\mathrm{raw}};\mathcal{B})
=
\frac{1}{|\mathcal{B}|}\sum_{b\in\mathcal{B}}
\ell\!\big(\Psi(\mathcal{G}_{\mathrm{eval}}(\theta_{\mathrm{sim}})a_b),y_b\big),
$$
where \(\mathcal{B}\) is the current minibatch and \(\mathcal{G}_{\mathrm{eval}}(\theta_{\mathrm{sim}})\) denotes the full optical forward map evaluated at the projected design. Under that formulation, each adjoint iteration depends on the sampled data. Once \(T^\star\) is fixed, however, the adjoint sources in the two-stage workflow depend only on the scattering residuals $
T(\theta_{\mathrm{sim}})-T^\star,
R(\theta_{\mathrm{sim}}),
$
rather than on labeled minibatches. As a result, the expensive inverse-design loop is batch-free, and the realization stage depends on operator matching rather than dataset size.

\subsection{Task evaluation}
Task metrics are computed only after the realized transmission matrix \(T(\theta_{\mathrm{sim}})\) has been extracted from simulation. For an all-optical setting,
$$
u_{\mathrm{eval}} = T(\theta_{\mathrm{sim}})a,
\qquad
I_{\mathrm{eval}} = |u_{\mathrm{eval}}|^2,
\qquad
g_{\mathrm{eval}} = I_{\mathrm{eval}}.
$$
With a fixed downstream evanescent stage, whose forward matrix is \(E\) (that is, \(O=E\)), and using the same trained linear readout \((W,b)\), we instead have
$$
u_{\mathrm{eval}} = E\,T(\theta_{\mathrm{sim}})\,a,
\qquad
I_{\mathrm{eval}} = |u_{\mathrm{eval}}|^2,
\qquad
g_{\mathrm{eval}} = W^\top I_{\mathrm{eval}} + b.
$$
The reported accuracy and cross-entropy are computed from the logits \(g_{\mathrm{eval}}\).

\subsection{Preprocessing, encoding, and locality}
To satisfy the available port budget, we preprocess the MedMNIST \cite{Yang2023MedMNISTv2} and RSSCN7 \cite{Zou2015RSSCN7} datasets before optical encoding. For MedMNIST, we flatten grayscale images, normalize them to \([0,1]\), and reduce them to sixteen dimensions by principal component analysis \cite{Jolliffe2002PCA},
$$
x=U_{\mathrm{PCA},r}^\top(z-\mu),
$$
so that the optical input dimension matches the available number of ports while preserving most of the sample variance. We apply an analogous dimensionality-reduction step to RSSCN7.

The study uses two optical encoding families,
$$
\phi_{\mathrm{amp}}(x)=x,
\qquad
\phi_{\mathrm{phase}}(x)=e^{-i\pi x}.
$$
The former is a real-amplitude encoding, and the latter is a phase-only encoding on the unit circle. Across the reported experiments, MedMNIST uses amplitude encoding, whereas RSSCN7 and the Yin-Yang dataset use phase encoding. Empirically, amplitude encoding performed better for MedMNIST, while phase encoding was more effective for the latter datasets.

To reduce the routing burden on the trainable device, we impose locality as an architectural prior during surrogate training. For a square \(N\times N\) matrix, a banded mask with half-bandwidth \(w\) reduces the number of free entries from \(N^2\) to
$$
N(2w+1)-w(w+1).
$$
Such masking discourages long-range direct routing inside a single freeform region and shortens the propagation length required for inverse design.

\begin{figure}[htbp]
\centering
\includegraphics[width=\linewidth]{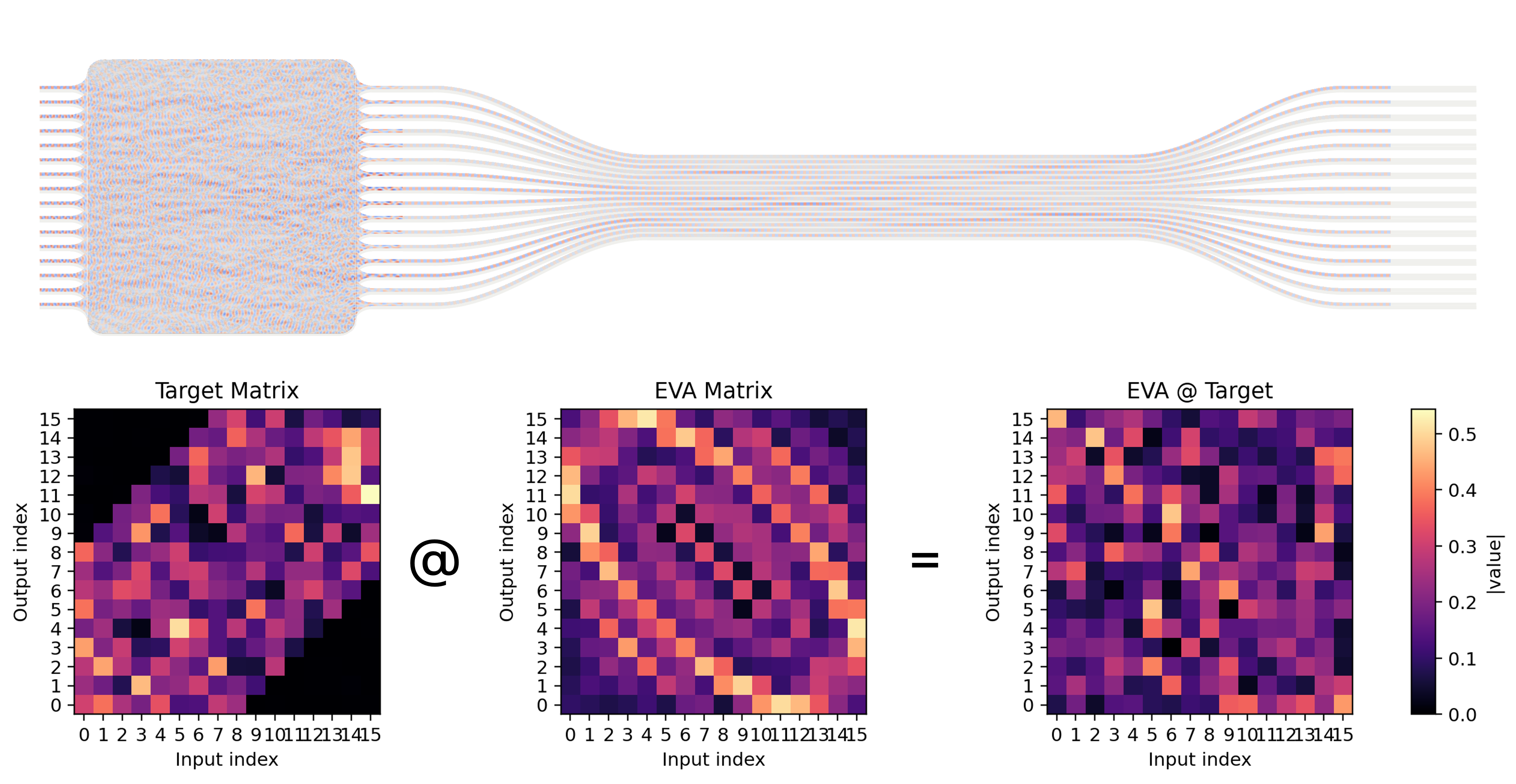}
\caption{Banded router combined with an evanescent-coupling region. (Top) Propagated field in the full device for a sample from the RSSCN7 dataset. (Bottom) The product of a banded target matrix (left) with a dense evanescent-coupling matrix (center) yields a fully dense effective operator (right), confirming that global mixing can be recovered from an individually sparse trainable factor.}
\label{fig:banded}
\end{figure}

Locality at one stage does not preclude global mixing after composition. If \(A\) and \(B\) are banded with half-bandwidths \(w_A\) and \(w_B\), then \(AB\) has half-bandwidth at most \(w_A+w_B\) \cite{HornJohnson2012MatrixAnalysis}. When the summed bandwidth spans the matrix, the product can therefore be fully dense even though each factor is individually banded. More broadly, structured sparse factorizations can represent or approximate dense operators efficiently \cite{Dao2019Butterfly}. For RSSCN7, the learned local router is banded, while dense mixing is completed by the fixed evanescent stage. Figure~\ref{fig:banded} illustrates this composition.

The three datasets studied here instantiate the same framework in different regimes. MedMNIST is an all-optical example with 16 optical inputs and 6 output classes. The local device itself acts as the classifier, there is no downstream optical chain, and the prediction is read directly from the output intensities. RSSCN7 is a chained example with 16 optical inputs, a learned \(16\times 16\) local router, a fixed \(16\times 16\) evanescent stage, and a linear readout on detected intensities. The Yin-Yang dataset is used to test the same framework on a nonlinear decision-boundary problem, where the optical front end is followed by intensity detection and a small digital readout.

\section{Results}

We evaluate the method on three tasks. MedMNIST is an all-optical classification problem in which the realized optical operator directly maps a 16-dimensional encoded feature vector to six class channels. RSSCN7 is a classification problem in which a learned \(16 \times 16\) local router is followed by a fixed \(16 \times 16\) evanescent-coupling stage. Together, these tasks probe complementary aspects of the method. MedMNIST tests whether a high-performing surrogate classifier can be transferred to an all-optical inverse-designed task with minimal performance degradation. RSSCN7 tests whether a local router can still be realized effectively when it is constrained to a banded matrix, with the evanescent stage providing the additional mixing needed to realize a dense effective operator. Finally, we study a nonlinear toy problem based on the Yin-Yang dataset to illustrate how the framework can support neuromorphic computing with digital readout.

For MedMNIST, the surrogate stage identifies a \(6 \times 16\) passive target matrix that reaches 98.75\% test accuracy with a test cross-entropy of 0.0887. The inverse-design stage then transfers this target to a device with a 25 nm design-region pixel size and a 200 nm conic-filter radius, corresponding to a minimum feature size of 400 nm. This feature size is large enough to be fabricated with maskless lithography or electron-beam lithography. The design region measures \(26.4 \times 31.35~\mu\mathrm{m}\), with effective indices \(n_{\mathrm{eff}}^{220\,\mathrm{nm}} = 2.86\) and \(n_{\mathrm{eff}}^{150\,\mathrm{nm}} = 2.55\).

\begin{figure}[htbp]
\centering
\safeincludegraphics[width=\linewidth]{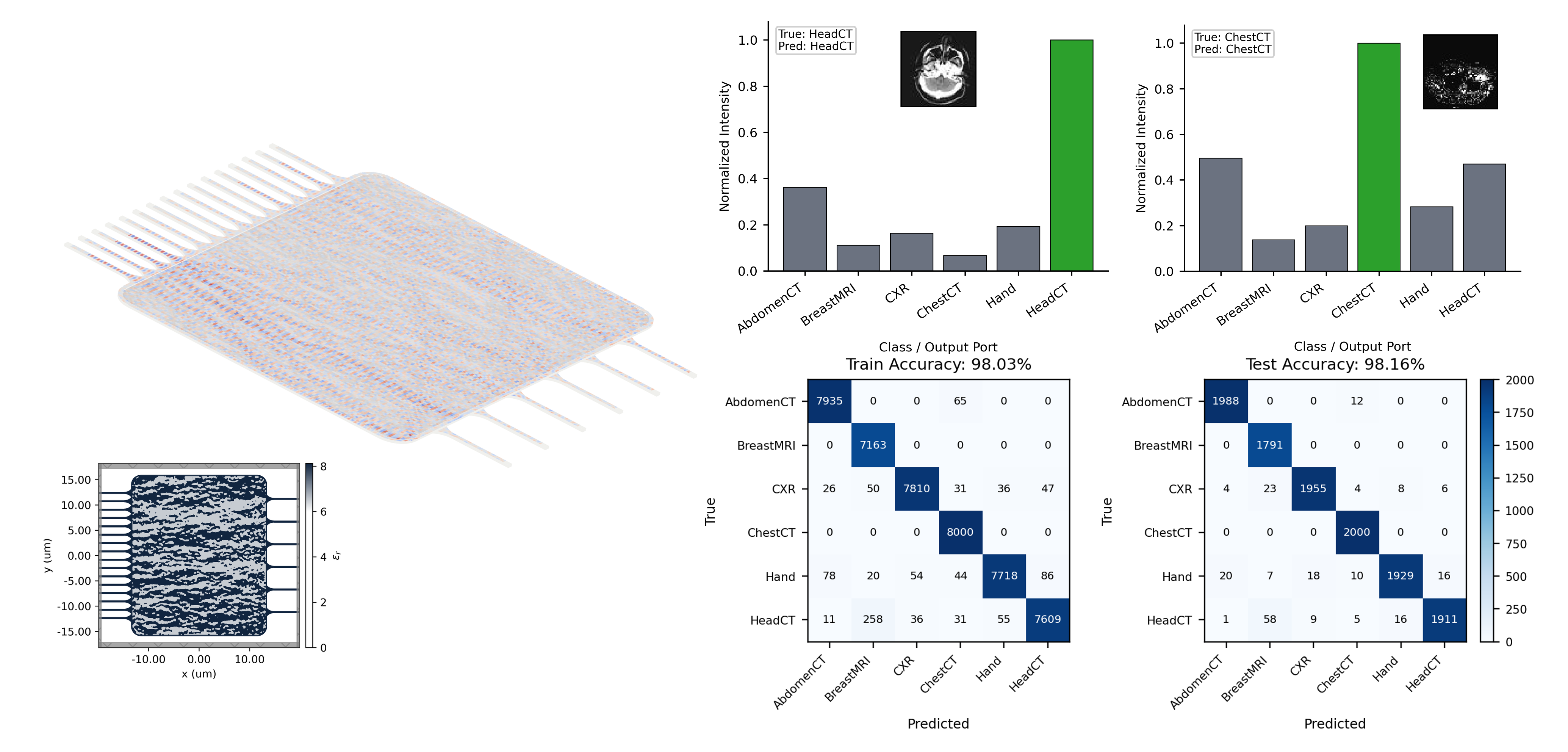}
\caption{Inverse-designed MedMNIST classifier. (Left) Propagated field for a sample of class HeadCT in the optimized design region. (Inset, bottom) Permittivity distribution of the freeform design region. (Top) Normalized output intensities for two representative samples, with the correct class highlighted. (Bottom) Confusion matrices on the training and test sets, reaching $98.16\%$ test accuracy.}
\label{fig:medmnist-device}
\end{figure}

\begin{figure}[htbp]
\centering
\includegraphics[width=\linewidth]{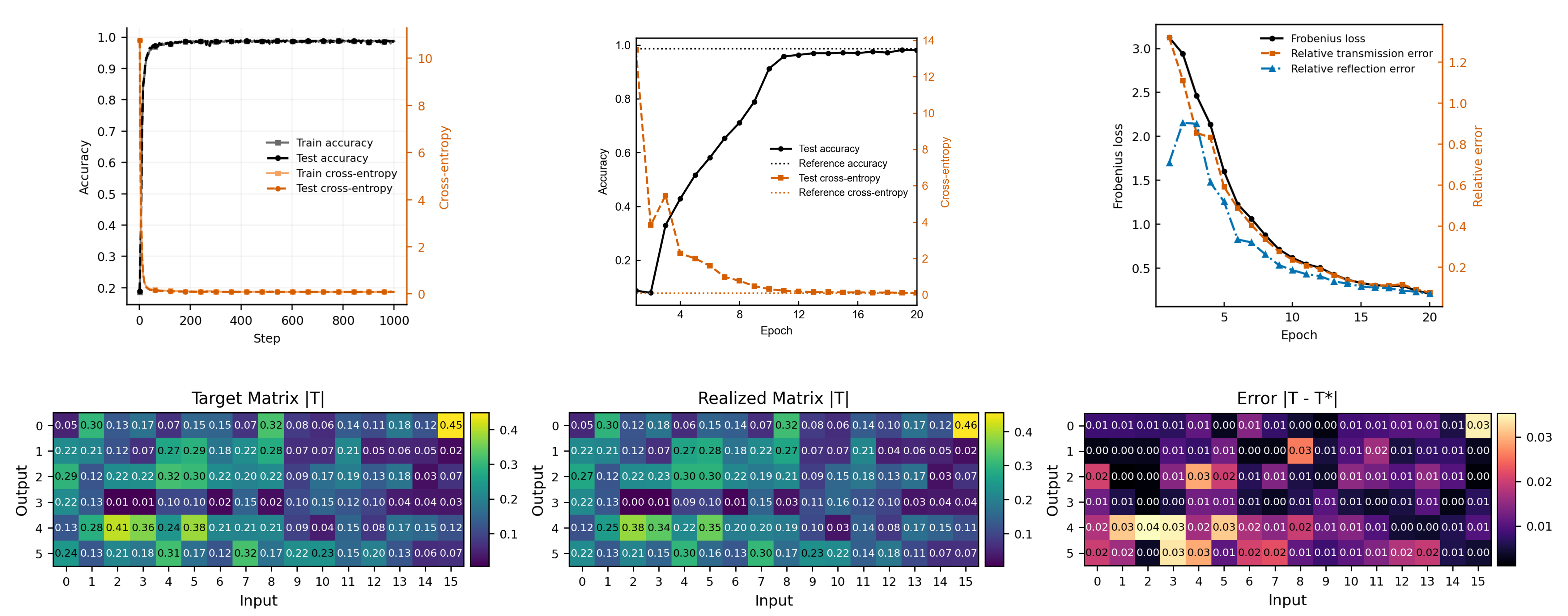}
\caption{Training dynamics and realized operator for the MedMNIST task. (Top, left to right) Surrogate training and test accuracy and cross-entropy loss over the surrogate-stage steps; inverse-design training and test accuracy and cross-entropy over 20 epochs; Frobenius-norm realization loss with relative transmission and reflection errors. (Bottom, left to right) Target surrogate matrix $T^{\star}$, simulated realized matrix $T(\theta_{\mathrm{sim}})$, and absolute error $|T(\theta_{\mathrm{sim}})-T^{\star}|$.}
\label{fig:medmnist-training}
\end{figure}

The realized device preserves the surrogate performance closely. After 20 optimization epochs, it reaches 98.16\% test accuracy and a test cross-entropy of 0.1133, only 0.59 percentage points below the selected surrogate target. At the same checkpoint, the normalized transmission residual is 0.0758, the reflection Frobenius norm is 0.0702, and the total optimization loss is 0.2088, including the fabrication-aware penalty. The inverse-designed structure has an insertion loss of 7.08 dB, compared with 6.80 dB for the selected target matrix. Overall, the adjoint method successfully designs a device that performs all-optical classification in only 20 optimization epochs with negligible accuracy loss. Compared with the MedNIST inverse-design study of Sved \emph{et al.}~\cite{Sved2026PNNAccelerators}, which reported 150 optimization epochs, the present workflow converges substantially faster. Representative field propagation, output intensities, and confusion matrices are shown in Fig.~\ref{fig:medmnist-device}, and the training dynamics and realized transmission matrix are shown in Fig.~\ref{fig:medmnist-training}.

For RSSCN7, the local router is optimized in the presence of a fixed evanescent-coupling region, followed by intensity measurement and digital readout. During surrogate optimization, we first obtain the S-parameters of the evanescent region and then combine them with the learned router through the Redheffer star product, i.e., the standard composition rule for cascaded scattering matrices, before intensity measurement and classification with a simple seven-neuron digital readout. The design region uses the same parameters as in the MedMNIST case. The evanescent region length is set to \(47.5~\mu\mathrm{m}\), chosen by sweeping the evanescence-only accuracy over lengths from \(40\) to \(50~\mu\mathrm{m}\) in \(0.5~\mu\mathrm{m}\) increments.

Despite the aggressive dimensionality reduction, the combined optical architecture improves substantially over simpler baselines. The surrogate training runs for 3000 epochs and yields a training accuracy of 55.27\%, a test accuracy of 52.14\%, and a cross-entropy loss of 1.1917. Because this is a heavily downsampled task, a linear ridge-classifier baseline reaches only 36.79\% test accuracy. Meanwhile, the bandwidth-constrained router alone and the evanescent-coupling region alone achieve 48.75\% and 42.5\%, respectively. The combined method therefore improves accuracy by 15.35 percentage points relative to the linear-classifier baseline.

\begin{figure}[htbp]
\centering
\safeincludegraphics[width=\linewidth]{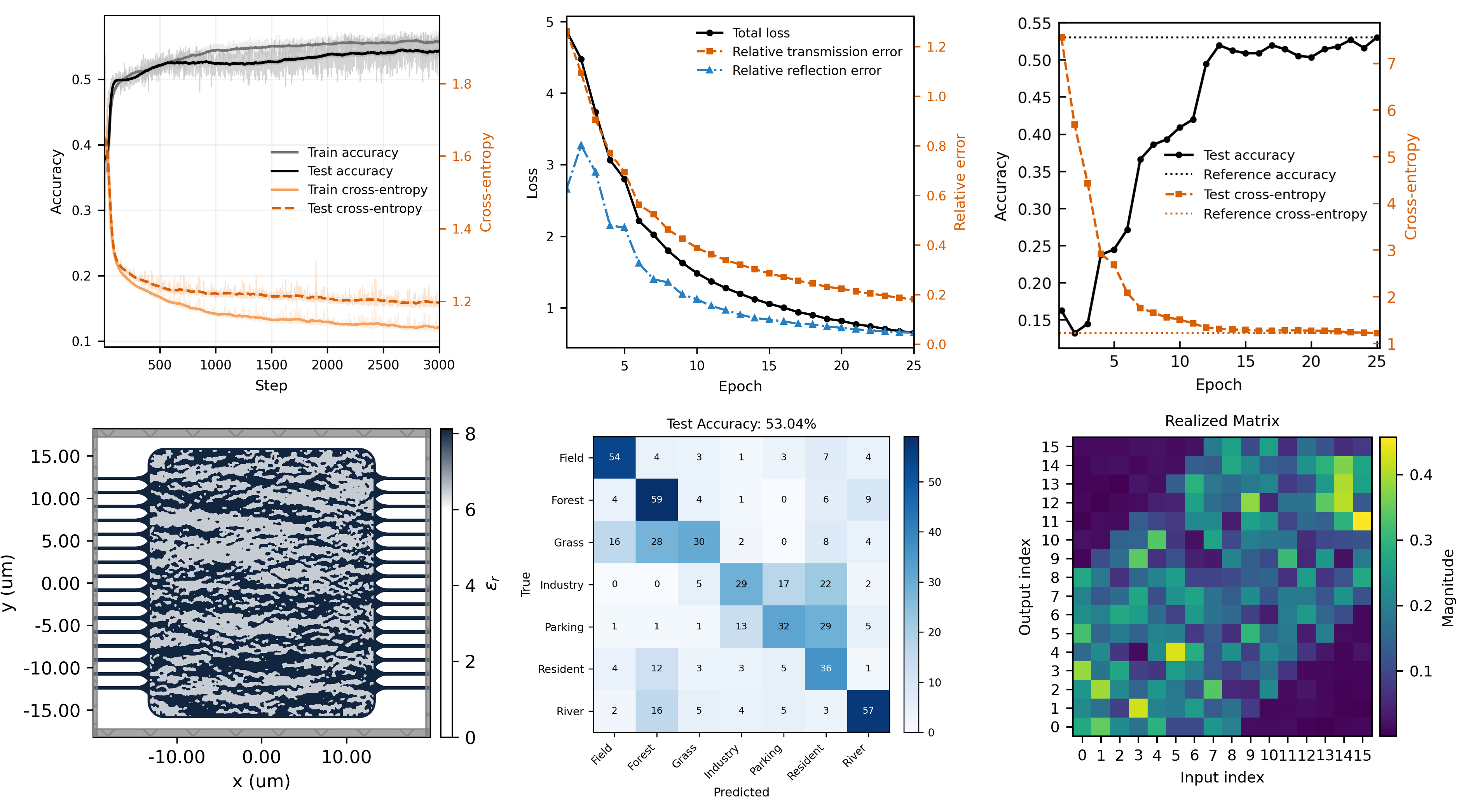}
\caption{RSSCN7 results. (Top, left to right) Surrogate training and test accuracy and cross-entropy over 3000 epochs; Frobenius realization loss with relative transmission and reflection errors over 25 inverse-design epochs; inverse-design training and test accuracy and cross-entropy. (Bottom, left) Permittivity distribution of the realized banded-router device. (Bottom, center) Confusion matrix on the RSSCN7 test set, reaching $53.04\%$ accuracy. (Bottom, right) Magnitude of the realized transmission matrix.}
\label{fig:rsscn7}
\end{figure}

The inverse-design stage improves steadily over 25 epochs and finishes slightly above the surrogate model. By epoch 25, the test accuracy reaches 53.04\% and the test cross-entropy falls to 1.218, compared with 16.25\% accuracy at initialization, which is close to the \(1/7\) chance level of 14.29\%. Over the same interval, the loss, relative transmission error, and reflection error drop from 4.868, 1.263, and 0.626 to 0.6537, and 0.1808 to 0.0463, respectively. The simulated transmission loss at epoch 25 is 3.04 dB. The reference surrogate matrix has an insertion loss of 1.52 dB, so the realized router remains lossier than its target. This could likely be improved by adding a transmission penalty. Nevertheless, the inverse-designed device successfully performs the RSSCN7 task while requiring only 25 optimization epochs. The training dynamics, realized permittivity, confusion matrix, and realized transmission matrix are shown in Fig.~\ref{fig:rsscn7}.

\FloatBarrier

\begin{figure}[!t]
\centering
\safeincludegraphics[width=0.7\linewidth]{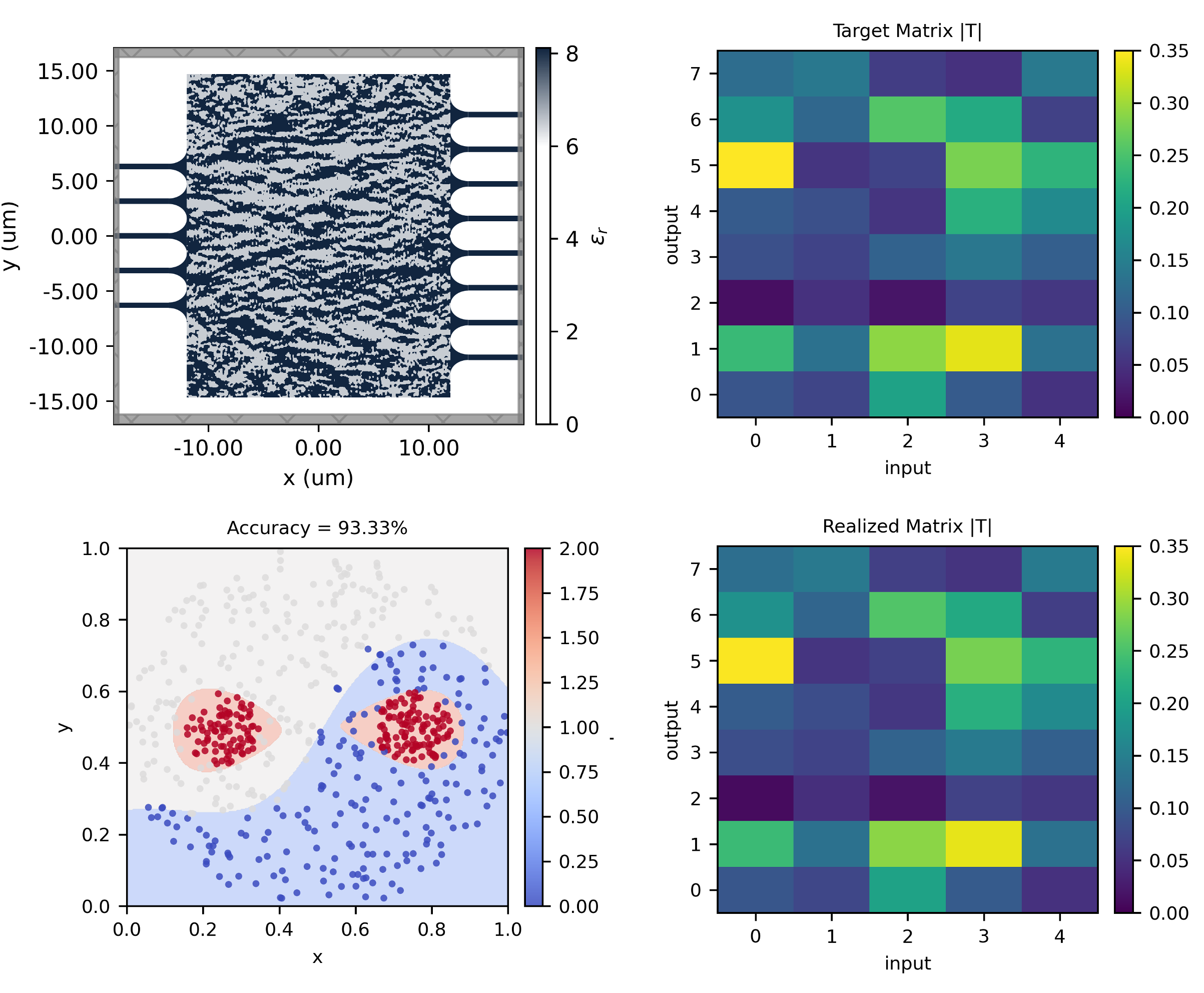}
\caption{Yin-Yang nonlinear task. (Top, left) Permittivity distribution of the realized device. (Top, right) Target transmission matrix from the surrogate stage. (Bottom, left) Decision boundary obtained from the realized device, reaching $93.67\%$ accuracy. (Bottom, right) Magnitude of the realized transmission matrix.}
\label{fig:yinyang}
\end{figure}

For the final nonlinear task based on the Yin-Yang dataset, the router uses five ports: four ports carry the encoded input and the fifth acts as a reference bias. The inputs are then projected onto eight output ports, corresponding to dimensionality expansion, before the intensities are measured and sent to a digital readout for classification. The linear-classifier baseline for this task is 65.83\%. Using only the evanescent region without routing increases the accuracy to 82.17\%, and adding the router raises the best surrogate-stage accuracy to 93.17\%. In the inverse-design stage, the design region measures \(24 \times 29~\mu\mathrm{m}\) with a uniform 25 nm design-grid size. The same filtering and projection scheme used in the previous tasks is applied here, and the optimization runs for 25 epochs, reaching an accuracy of 93.67\% at the end of the optimization. The realized permittivity, decision boundary, and target and realized transmission matrices are shown in Fig.~\ref{fig:yinyang}.

\section{Discussion}
The two-stage workflow reduces the number of full-wave simulations needed to train an inverse-designed photonic classifier. In a direct geometry-to-task pipeline, each adjoint epoch requires one forward and one backward solve per sample in the minibatch. The simulation count scales as $O(N_{\mathrm{fwd}}\cdot N_{\mathrm{bwd}}\cdot B)$, where $B$ is the batch size. Once $T^{\star}$ is fixed, the adjoint sources depend only on the transmission and reflection residuals. The loop collapses to $O(N_{\mathrm{fwd}}\cdot N_{\mathrm{bwd}})$ per epoch and becomes independent of dataset size. For our datasets, with $B$ between a few hundred and a few thousand samples, this is a simulation-budget reduction of one to three orders of magnitude. Both MedMNIST and RSSCN7 realization stages converge in 20 to 25 epochs, compared with $\sim$150 epochs in comparable end-to-end studies \cite{Sved2026PNNAccelerators}. This is a structural property of decoupling label fitting from electromagnetic realization, not a numerical speedup specific to our implementation.

A second benefit is the smoothness of the realization loss. The Frobenius residual specifies the required amplitude and phase at each output port, so the adjoint sources depend only on modal scattering quantities and not on labels or minibatch statistics. This avoids the difficulty of optimizing intensity-domain cross-entropy, where the $|u|^{2}$ nonlinearity introduces phase ambiguity in the gradient and produces flat regions in the loss landscape. We observe monotone decay of the transmission residual in all three tasks. The realized accuracy matches or slightly exceeds the surrogate accuracy once the residual falls below 0.2.

The banded-router plus evanescent-stage architecture addresses a second bottleneck that grows with the number of ports. A fully local freeform router must connect port 0 to port $N-1$, which requires a propagation length of order $N$ times the port pitch and forces the design region to be long. A banded structure with half-bandwidth $w$ limits the direct routing distance to $w$ ports. The fixed evanescent region supplies the remaining mixing. Since the product of two banded matrices has half-bandwidth at most $w_A+w_B$, the combined operator can be fully dense. For the 16-port RSSCN7 task, a banded router with $w\approx N/2$ and a $47.5~\mu\mathrm{m}$ evanescent stage reproduces a dense effective operator in a design region half as long as a fully local router would need. This halves the number of design grid points simulated per adjoint iteration.

Our study has several limitations. All simulations use a two-dimensional effective-index model at $\lambda=1.55~\mu\mathrm{m}$. This is a standard and experimentally validated reduction for slab waveguide geometries at the relevant index contrast \cite{Nikkhah2024VectorMatrix}, but it does not capture out-of-plane radiation or wavelength dispersion.

Loss is another limitation. The MedMNIST device has $7.08$~dB insertion loss, and the RSSCN7 router shows $1.5$~dB excess loss over its target. A dedicated transmission penalty, or tighter bounds on the singular-value vector \(s\), could improve these numbers.

The absolute RSSCN7 accuracy is also modest because the input is compressed to 16 principal components, for which the linear baseline is only $36.79\%$. The more relevant figure of merit is the $15.35$ percentage-point gain over that baseline. Finally, our results are simulation-based. Experimental validation on silicon photonics is the natural next step.

The workflow is agnostic to the choice of full-wave solver and to the fabrication platform. It extends to 3D FDTD simulation, broadband objectives, and material systems such as silicon nitride or thin-film lithium niobate. It can also be applied by cascading multiple surrogate-designed blocks with fixed passive stages, enabling deeper effective networks within the same simulation budget. The observation that a matrix-level surrogate can replace an entire electromagnetic realization loop during task training suggests a route to hardware-in-the-loop co-design, where only the final matching step requires the solver.

\section{Conclusion}
We introduced a surrogate-guided inverse-design framework for compact photonic neural networks. Task training is performed on a passive complex matrix with bounded singular values. Electromagnetic realization is then posed as a batch-free adjoint problem driven by a Frobenius-norm transmission residual. The workflow reduces the per-epoch simulation cost from $O(N_{\mathrm{fwd}}\cdot N_{\mathrm{bwd}}\cdot B)$ to $O(N_{\mathrm{fwd}}\cdot N_{\mathrm{bwd}})$ and yields a smoother loss than intensity-domain objectives. A banded router mixed by a fixed evanescent-coupling region halves the required design region while realizing dense effective operators. We validated the framework on three tasks. The MedMNIST all-optical classifier matches its surrogate accuracy within $0.6$ percentage points after 20 adjoint epochs. The RSSCN7 router improves on a linear readout baseline by more than 15 percentage points. A Yin-Yang task confirms support for nonlinear decision boundaries. These results establish surrogate-guided inverse design as a practical route to training compact photonic processors with simulation budgets orders of magnitude smaller than direct geometry-to-task pipelines. Experimental validation on silicon photonics is the natural next step.

\begin{backmatter}
\bmsection{Funding}
This work was supported by the Optica Challenge Award.
\bmsection{Acknowledgment}
The authors acknowledge Flexcompute for support through the Tidy3D educational license program. Electromagnetic simulations were performed using the Tidy3D finite-difference time-domain solver.

\bmsection{Disclosures}
The authors declare no conflicts of interest.

\bmsection{Data availability}
Data underlying the results presented in this paper are not publicly available at this time but may be obtained from the authors upon reasonable request.
\end{backmatter}

\bibliography{sample}

\end{document}